\documentclass [aps,pra,twocolumn,superscriptaddress]{revtex4}
\usepackage{amsmath}
\usepackage{graphicx}
\usepackage{color}

\begin{document}

\title{How nonlocal damping reduces plasmon-enhanced fluorescence in ultranarrow gaps}
\author{C. Tserkezis}
\affiliation{Department of Photonics Engineering, Technical University of Denmark, {\O}rsteds Plads 343, DK-2800 Kgs. Lyngby, Denmark}
\email{ctse@fotonik.dtu.dk}
\author{N. Asger Mortensen}
\affiliation{Department of Photonics Engineering, Technical University of Denmark, {\O}rsteds Plads 343, DK-2800 Kgs. Lyngby, Denmark}
\affiliation{Center for Nanostructured Graphene, Technical University of Denmark, {\O}rsteds Plads 343, DK-2800 Kgs. Lyngby, Denmark}
\affiliation{Center for Nano Optics, University of Southern Denmark, Campusvej 55, DK-5230 Odense M, Denmark}
\author{Martijn Wubs}
\affiliation{Department of Photonics Engineering, Technical University of Denmark, {\O}rsteds Plads 343, DK-2800 Kgs. Lyngby, Denmark}
\affiliation{Center for Nanostructured Graphene, Technical University of Denmark, {\O}rsteds Plads 343, DK-2800 Kgs. Lyngby, Denmark}

\begin{abstract}
The nonclassical modification of plasmon-assisted fluorescence enhancement is theoretically explored by placing two-level dipole emitters at the narrow gaps encountered in canonical plasmonic architectures, namely dimers and trimers of different metallic nanoparticles. Through detailed simulations, in comparison with appropriate analytical modelling, it is shown that within classical electrodynamics, and for the reduced separations explored here, fluorescence enhancement factors of the order of $10^{5}$ can be achieved, with a divergent behaviour as the particle touching regime is approached. This remarkable prediction is mainly governed by the dramatic increase in excitation rate triggered by the corresponding field enhancement inside the gaps. Nevertheless, once nonclassical corrections are included, the amplification factors decrease by up to two orders of magnitude and a saturation regime for narrower gaps is reached. These nonclassical limitations are demonstrated by simulations based on the generalised nonlocal optical response theory, which accounts in an efficient way not only for nonlocal screening, but also for the enhanced Landau damping near the metal surface. A simple strategy to introduce nonlocal corrections to the analytic solutions is also proposed. It is therefore shown that the nonlocal optical response of the metal imposes more realistic, finite upper bounds to the enhancement feasible with ultrasmall plasmonic cavities, thus providing a theoretical description closer to state of the art experiments.
\end{abstract}
\maketitle


\section{Introduction}\label{Sec:intro}

Plasmonic nanostructures are widely explored for improving fluorescence of organic molecules or quantum dots \cite{fu_lpr3,ming_jpcl3,chen_rprogph75,choudhury_jpcl4,deng_pccp15,darvill_pccp15,bauch_plasm9,bozhevolnyi_optica3}, owing their popularity mainly to their unique ability to focus and enhance electromagnetic fields at the nanoscale \cite{xu_pre62,zayats_jopta5,bharadwaj_aop1,neacsu_nl10,schuller_natmat9}. Several geometries have been explored over the years for studying and optimizing emission properties of two-level systems, including flat metal surfaces \cite{drexhage_jum1,chance_acp37,shimizu_prl89}, single metallic nanoparticles \cite{gersten_jcp75,ruppin_jcp76,anger_prl96,kuhn_prl97}, and aggregates thereof \cite{muskens_nl7,bek_nl8,kinkhabwala_natphot3}. In general, plasmonic nanocavities are beneficial for such studies, as they provide significantly faster spontaneous emission rates \cite{russell_natphot6} and tremendous Purcell \cite{akselrod_natphot8} and fluorescence enhancement factors \cite{schmelzeisen_nn4,rose_nl14,puchkova_nl15}. Such cavities are also exploited for single-photon emission \cite{esteban_prl104}, while they have recently led to strong emitter-plasmon coupling at the single-molecule level \cite{santhosh_natcom7,chikkaraddy_nat535}. Unlike single plasmonic nanoparticles, for which emitters placed in close proximity to the metal couple to dark higher-order modes resulting in fluorescence quenching \cite{anger_prl96}, in plasmonic nanogaps this quenching is strongly suppressed due to modification of the hybrid plasmon modes and their coupling with the emitter \cite{jun_prb78,faggiani_acsphot2,kongsuwan_arxiv}, thus placing them at the forefront of efforts to efficiently control emission dynamics.

Fluorescence enhancement and emitter coupling to plasmonic nanocavities are commonly studied theoretically within the framework of classical electrodynamics, which is often adequate to provide a good quantitative agreement with experiments \citep{anger_prl96}. Departing from this description is more common when one is interested in the dynamics of quantum emitter coupling with plasmonic nanostructures \cite{manjavacas_nl11,gonzalez_prl110,delga_prl112,yang_prb95}. Nevertheless, the tremendous recent advances in cavity minimisation have now led to a necessity for inclusion of nonclassical effects such as nonlocal screening and electron spill-out \cite{abajo_jpcc112,mcmahon_prl103,raza_jpcm27,zhu_natcom7} in the modelling of the plasmonic nanostructures \cite{filter_ol39,li_ieeeph8}. Eearlier theoretical studies based on the hydrodynamic Drude model, which accounts well for nonlocal blueshifts in noble metals, have already predicted an impact of nonlocality on emitter-plasmon coupling \cite{leung_prb42,hider_prb66,moroz_annphys315,ren_prb94}. To advance one step further, we have recently explored fluorescence near single nonlocal plasmonic particles by implementing the generalised nonlocal optical response (GNOR) theory \cite{mortensen_natcom5}, which accounts in addition also for surface-enhanced Landau damping \cite{uskov_plas9,vagov_prb93,shahbazyan_prb94}. In that case, a significant decrease in fluorescence enhancement for emitters coupled to individual homogeneous noble-metal nanospheres or nanoshells has to be anticipated \cite{tserkezis_nscale8}. Since nanocavities behave substantially differently from isolated particles as far as their coupling with emitters is concerned \cite{chikkaraddy_nat535,sun_apl98}, it is important to explore the influence of nonlocality and plasmon damping, as predicted by the GNOR model, in such situations as well.

Fluorescent molecules and quantum dots, modelled as classical electric dipoles, are coupled here with some of the canonical plasmonic architectures frequently encountered both in theory and experiments. In particular, we place emitters at the gaps formed by metallic nanosphere and nanoshell dimers, bowtie antennas, and chains of three nanospheres either identical or in the configuration of a self-similar nanolens \cite{li_prl91,almpanis_ol37}. In all situations, calculations within the local response approximation (LRA) of classical electrodynamics predict a dramatic fluorescence enhancement that can approach $10^{6}$, as a result of the unprecedented increase in the near field responsible for the excitation of the molecule in the plasmonic cavity and the improved radiative properties of nearly-touching nanoparticle systems. By subsequently adopting a nonlocal description of the metallic component, we show that the increased plasmon damping, intrinsic in GNOR, leads to a corresponding reduction of fluorescence enhancement by up to two orders of magnitude. The divergent enhancement predicted by LRA for decreasing gap widths tends to saturate within GNOR, although further studies fully accounting for electron spill-out are required as a more conclusive step to verify this behaviour for even narrower gaps. In addition to the numerical calculations we also apply analytic solutions for the field enhancement and emission in the nanocavities \cite{sun_apl97,sun_apl98}, which we modify here to include nonlocal effects in an efficient way. Our work thus shows that the nonlocal optical response of metals imposes additional, more realistic upper bounds to the fluorescence enhancement achievable through plasmonic nanocavities \cite{bozhevolnyi_optica3}, and becomes important when few- or sub-nm gaps are considered.

\section{Theoretical background}\label{Sec:methods}

Numerical simulations are performed with a commercial finite-element method (COMSOL Multiphysics 5.0, RF module) \cite{comsol}, appropriately adapted to include nonlocal effects \cite{nanoplorg}. All architectures considered are embedded in air, which is described by a dielectric constant $\varepsilon = 1$. For the LRA simulations the plasmonic nanostructures are described by the experimental dielectric function ($\varepsilon_{\mathrm{exp}}$) of silver, as measured by Johnson and Christy \cite{johnson_prb6}. In the nonlocal simulations the metal dielectric function follows a Drude model, $\varepsilon_{\mathrm{m}} = \varepsilon_{\infty} - \omega_{\mathrm{p}}^{2}/[\omega (\omega + \mathrm{i} \gamma]]$, where $\omega$ is the angular frequency of light, $\omega_{\mathrm{p}}$ is the plasma frequency of silver, $\gamma$ is the damping rate and $\varepsilon_{\infty}$ the contribution of core electrons, calculated here by subtracting the Drude part from $\varepsilon_{\mathrm{exp}}$. In all simulations we use $\hbar \omega_{\mathrm{p}} =$ 8.99 eV and $\hbar \gamma =$ 0.025 eV \cite{raza_jpcm27}. Within GNOR, one solves numerically the system of coupled equations \cite{mortensen_natcom5}
\begin{equation}\label{Eq:NonlocalWave}
\begin{split}
& \nabla \times \nabla \times \mathbf{E} (\mathbf{r}, \omega) = \left(\frac{\omega}{c}\right)^{2} \varepsilon_{\infty} \mathbf{E} (\mathbf{r}, \omega) + \mathrm{i} \omega \mu_{0} \mathbf{J} (\mathbf{r}, \omega)\\
& \left[\frac{\beta^{2}}{\omega \left(\omega + \mathrm{i} \gamma\right)} + \frac{D}{\mathrm{i} \omega} \right] \nabla \left[\nabla \cdot \mathbf{J} (\mathrm{r}, \omega) \right] + \mathbf{J} (\mathrm{r}, \omega) = \sigma \mathbf{E} (\mathrm{r}, \omega)~,
\end{split}
\end{equation}
where $\mathbf{E} (\mathbf{r}, \omega)$ and $\mathbf{J} (\mathbf{r}, \omega)$ are the (position $\mathbf{r}$-dependent) electric field and induced current density, respectively; $\sigma = \mathrm{i} \varepsilon_{0} \omega_{\mathrm{p}}^{2}/(\omega + \mathrm{i} \gamma)$ is the Drude conductivity, and $\varepsilon_{0}$ and $\mu_{0}$ are the vacuum permittivity and permeability respectively, related to the velocity of light in vacuum through $c = 1/\sqrt{\varepsilon_{0} \mu_{0}}$. The hydrodynamic parameter $\beta$ is taken equal to $\sqrt{3/5} \; v_{\mathrm{F}}$, where $v_{\mathrm{F}} = 1.39 \cdot 10^{6}$ m s$^{-1}$ is the Fermi velocity of silver \cite{raza_jpcm27}, while for the diffusion constant $D$ we use $D = 2.684 \cdot 10^{4}$ m$^{2}$ s$^{-1}$ \cite{tserkezis_scirep6}. As can be seen from Eqs.~(\ref{Eq:NonlocalWave}), GNOR constitutes an extension of the traditional hydrodynamic model for nonlocal plasmonics \cite{dasgupta_prb24,fuchs_prb35,ruppin_prb45,drachev_jetp94,chang_sscom133}, with the addition of a diffusion term that accounts in a semi-classical way for surface-enhanced Landau damping. We note that a diffusive-like term originating from the bulk is already included in the standard hydrodynamic model, but at optical frequencies it is practically negligible \cite{tserkezis_ijmpb}. In this respect, GNOR can be directly introduced into any analytical or numerical implementation of the hydrodynamic model, simply by setting $\beta^{2} \rightarrow \beta^{2} + D (\gamma - \mathrm{i} \omega)$.

Throughout the paper emitters are placed at the middle of gaps of width $d$ formed between silver nanoparticles, as shown for example in the inset of Fig.~\ref{fig1}(a). The emitters are modelled as classical electric dipoles, with their dipole moment $\mathbf{p}_{\mathrm{d}}$ parallel to the dimer axis. A plane wave with electric field $\mathbf{E}_{0}$ polarised along the dimer axis excites the two-level system from the ground state to its excited state at wavelength $\lambda$. The emitter is then assumed to decay back to the ground state by emitting a photon of the same wavelength. Fluorescence ($\gamma_{\mathrm{em}}$) is described as the combined result of two independent procedures: emitter excitation, described by a rate $\gamma_{\mathrm{exc}}$, and emission into the environment where the energy is either radiated or absorbed by the particle, described by the quantum yield $q$ \cite{anger_prl96}. The total enhancement is then obtained from
\begin{equation}
\frac{\gamma_{\mathrm{em}}}{\gamma_{\mathrm{em}}^{0}} = \frac{\gamma_{\mathrm{exc}}}{\gamma_{\mathrm{exc}}^{0}} \frac{q}{q^{0}}~,
\end{equation}
where the superscript ``0'' denotes the corresponding values in the absence of a plasmonic environment. The excitation rate is calculated through the electric field at the position $\mathbf{r}_{\mathrm{d}}$ of the emitter,
\begin{equation}
\gamma_{\mathrm{exc}} \propto \left| \mathbf{p}_{\mathrm{d}} \cdot \mathbf{E} (\mathbf{r}_{\mathrm{d}}) \right|^{2}~,
\end{equation}
while the quantum yield is obtained from the ratio of radiative ($\gamma_{\mathrm{r}}$) to total decay rate,
\begin{equation}
q = \frac{\gamma_{\mathrm{r}}/\gamma_{\mathrm{r}}^{0}}{\gamma_{\mathrm{r}}/\gamma_{\mathrm{r}}^{0} + \gamma_{\mathrm{abs}}/\gamma_{\mathrm{r}}^{0}}~.
\end{equation}
Here, $\gamma_{\mathrm{abs}}$ is the absorptive decay rate, and the different decay rate ratios are obtained through the corresponding power in the absence (radiated energy) or presence (both radiated and absorbed energy) of a plasmonic environment. We have further assumed that the intrinsic emitter quantum yield $q^{0} = 1$ \cite{anger_prl96}. Finally, extinction cross sections ($\sigma_{\mathrm{ext}}$) are obtained by adding the corresponding scattering and absorption cross sections, which are calculated through the power radiated to the far field and the total loss at the particles, respectively \cite{comsol}.

We complement our numerical simulations by an approximate analytic model based on coupled-mode theory, originally developed by Sun and Khurgin \cite{sun_apl98}, which we adapt here to include nonlocal effects and employ to study nanosphere and nanoshell dimers. According to this model, the total enhancement can be obtained as the steady-state solution of a coupled-mode approach which contains approximations for the radiative and nonradiative decay rates of the individual nanoparticles, and the corresponding plasmon resonance frequencies $\omega_{\ell}$ for all modes of order $\ell$ in a spherical-wave multipole expansion. The radiative decay rate of a nanosphere of radius $R$ is approximated by $\gamma_{\mathrm{r}} = 16 \pi^{3} \omega R^{3}/(3 \varepsilon \lambda^{3})$. Similarly, the nonradiative (absorptive) decay rate is estimated as $\gamma_{\mathrm{nr}} = \omega \mathrm{Im}\varepsilon_{\mathrm{m}}/\mathrm{Re}\varepsilon_{\mathrm{m}}$. In our analytic calculations $\varepsilon_{\mathrm{m}}$ is given by a Drude model with $\varepsilon_{\infty} = 4.55$, $\hbar \omega_{\mathrm{p}} =$ 9 eV and $\hbar \gamma =$ 0.125 eV, values for which both the far-field spectra and the distance dependence of fluorescence enhancement near a single Ag nanosphere with $R =$ 20 nm agree excellently with the corresponding results obtained when the metal is described by $\varepsilon_{\mathrm{exp}}$. Taking the asymptotic form of the spherical Bessel and Hankel functions \cite{Arfken_Academic2000} involved in the scattering matrix of a single sphere as obtained within Mie theory \cite{Bohren_Wiley1983}, the plasmon modes are found to follow \cite{christensen_nn8,raza_natcom6}
\begin{equation}\label{Eq:SphereLSPsLocal}
\omega_{\ell} = \omega_{\mathrm{p}} \sqrt{\frac{\ell}{\ell \varepsilon_{\infty} + \left(\ell + 1\right) \varepsilon}}~.
\end{equation}

Nonlocal corrections can be efficiently introduced into the coupled-mode model by adding to the plasmon resonance solution the well-known nonlocal blueshift \cite{mcmahon_prl103,raza_jpcm27}. Using again asymptotic expressions for the Bessel functions in the Mie solution for the scattering matrix of a nonlocal metallic nanosphere \cite{raza_oex21,tserkezis_scirep6}, the resonance frequencies of the individual particles are modified to
\begin{equation}\label{Eq:SphereLSPsNonlocal}
\omega_{\ell} = \omega_{\mathrm{p}} \sqrt{\frac{\ell}{\ell \varepsilon_{\infty} + \left(\ell + 1\right) \varepsilon + 2\varepsilon \delta_{\mathrm{NL}}}}~.
\end{equation}
In the above equation, the nonlocal correction is given by $\delta_{\mathrm{NL}} = j_{\ell} (k_{\mathrm{L}} R) (\varepsilon - \varepsilon_{\infty})/[k_{\mathrm{L}}R j_{\ell}^{'} (k_{\mathrm{L}}R) \varepsilon_{\infty}]$, where $j_{\ell}$ stands for the spherical Bessel function of order $\ell$, prime denotes derivative with respect to the argument, and $k_{\mathrm{L}}$ is the longitudinal wavenumber in the metal \cite{tserkezis_scirep6}. A further approximation for this relation can be found in Yan \emph{et al}. \cite{yan_prb88}. Finally, in order to capture the additional damping included in the GNOR model, we modify $\gamma_{\mathrm{nr}}$ by adding a size-dependent term, $v_{\mathrm{F}}/R$. This is exactly the phenomenological size-dependent damping correction adopted by Kreibig \emph{et al.} \cite{kreibig_zphys231,kreibig_surfsci156}, which GNOR reproduces within a more physical description \cite{mortensen_natcom5}.

The coupled-mode solution can be easily extended to describe a larger number of nanospheres and emitters through additional coupling terms \cite{sun_apl98}. On the other hand, in order to describe different kinds of nanoparticles, analytic solutions for their plasmon modes are required. Here we provide such a solution for the case of metallic nanoshells. A simple formula for the hybrid plasmon modes of this particle is available in literature \cite{prodan_sci302}, based however on the assumption that $\varepsilon$, $\varepsilon_{\infty}$, and $\varepsilon_{1}$ (the dielectric constant of the nanoshell core) are all equal to unity. To extend this we resort again to the analytic Mie solution for the scattering matrix of a core-shell particle \cite{tserkezis_nscale8,Bohren_Wiley1983} and apply the appropriate asymptotic expressions, arriving to
\begin{equation}\label{Eq:ShellLSPsNonlocal}
\ell \varepsilon_{\mathrm{m}} + \left(\ell + 1\right) \varepsilon = \left(\frac{R_{1}}{R}\right)^{2\ell +1} \frac{\ell \left(\ell + 1\right) \left(\varepsilon_{1} - \varepsilon_{\mathrm{m}}\right) \left(\varepsilon - \varepsilon_{\mathrm{m}}\right)}{\left(\ell + 1\right) \varepsilon_{\mathrm{m}} + \ell \varepsilon_{1}}~,
\end{equation}
where $R_{1}$ is the inner radius of the nanoshell. Solving Eq.~(\ref{Eq:ShellLSPsNonlocal}), which is quadratic with respect to $\varepsilon_{\mathrm{m}}$ (thus leading to two roots, $\varepsilon_{\mathrm{m}\pm}$), and then replacing $\varepsilon_{\mathrm{m}}$ by its Drude expression (setting $\gamma = 0$), gives two solutions,
\begin{equation}\label{Eq:ShellLSPsNonlocal2}
\omega_{\ell \pm} = \omega_{\mathrm{p}} \sqrt{\frac{1}{\varepsilon_{\infty} - \varepsilon_{\mathrm{m}\pm}}}~,
\end{equation}
one for the bonding, particle-like hybrid modes (at lower frequencies, ``-'' solution) and one for the antibonding, cavity-like ones (at higher frequencies, ``+'' solution) \cite{prodan_sci302}. It is straightforward to see that Eq.~(\ref{Eq:ShellLSPsNonlocal}) has the correct asymptotic form for $R_{1}/R \to 0$, as it leads to the well-known condition $\ell \varepsilon_{\mathrm{m}} + \left(\ell + 1\right) \varepsilon = 0$ \cite{raza_natcom6}, from which Eq.~(\ref{Eq:SphereLSPsLocal}) is derived. Repeating the same analysis for a nonlocal metallic nanoshell is a formidable task, since the analytic Mie solution is too lengthy (see for example the supplementary material in \cite{tserkezis_nscale8}). Nevertheless, if one is only interested in the particle-like modes, a simple correction similar to that of Eq.~(\ref{Eq:SphereLSPsNonlocal}) can be sufficient. Indeed, using
\begin{equation}\label{Eq:ShellLSPsNonlocal3}
\omega_{\ell -} = \omega_{\mathrm{p}} \sqrt{\frac{1}{\varepsilon_{\infty} - \varepsilon_{\mathrm{m}-} + 2\varepsilon \delta_{\mathrm{NL}}}}~,
\end{equation}
where $\delta_{\mathrm{NL}}$ is calculated as if the particle were a homogeneous metallic sphere of radius $R$, reproduces well the modal blueshifts. Here, we also correct the nonradiative decay rate to $\gamma_{\mathrm{nr}} = [1-(R_{1}/R)^{3}] \omega \mathrm{Im}\varepsilon_{\mathrm{m}}/\mathrm{Re}\varepsilon_{\mathrm{m}}$ (and with $v_{\mathrm{F}}/R$ added on the right-hand side in the nonlocal case), to account for the reduction of losses due to the presence of a smaller quantity of metal. 

\section{Results and discussion}\label{Sec:results}

\begin{figure}[h]
\centerline{\includegraphics*[width=1\columnwidth]{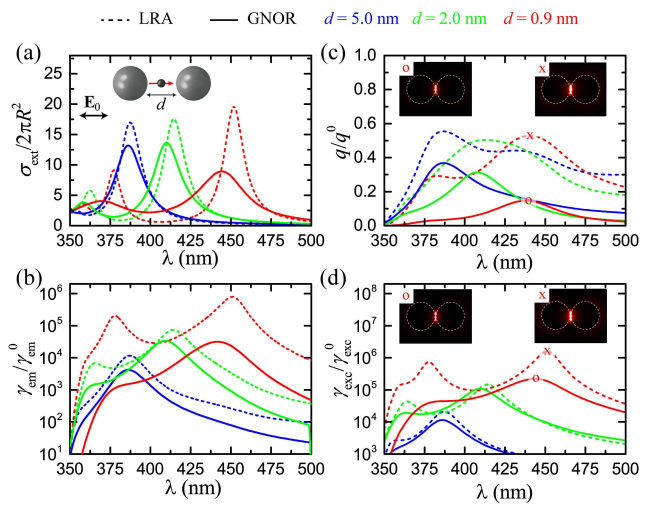}}
\caption{(a) Normalised extinction cross section ($\sigma_{\mathrm{ext}}$) of the Ag nanosphere dimer shown schematically in the inset, in the absence of an emitter. The two spheres (of equal radius $R =$ 20 nm) are separated by a gap of width $d$ and are excited by a plane wave with its electric field $\mathbf{E}_{0}$ parallel to the dimer axis. (b) Fluorescence enhancement ($\gamma_{\mathrm{em}}/ \gamma_{\mathrm{em}}^{0}$) spectra for a dipole emitter placed at the middle of the interparticle gap of the dimers in (a), with its dipole moment oscillating parallel to the dimer axis. (c) Normalised quantum yield ($q/q^{0}$) and (d) excitation rate enhancement ($\gamma_{\mathrm{exc}}/ \gamma_{\mathrm{exc}}^{0}$) spectra for the emitter of (b). The insets in (c) and (d) show electric field contours on resonance (saturated at a maximum enhancement of 200) around the dimer within LRA (right-hand insets) and GNOR (left-hand insets) upon (c) dipole emission and (d) excitation. In all cases blue, green, and red lines correspond to $d =$ 5, 2, and 0.9 nm respectively. Solid and dashed lines represent calculations with the GNOR and LRA models, respectively.}\label{fig1}
\end{figure}

As a first, typical example of a narrow plasmonic cavity, we consider the commonly encountered nanosphere dimer, and study fluorescence enhancement for an emitter placed at the middle of its gap. The two Ag nanospheres have a radius $R =$ 20 nm, a size which allows direct comparison with the case of an emitter close to a single Ag nanosphere \cite{tserkezis_nscale8}, and it immediately displays the fundamental differences between the two configurations. The far-field response of this dimer in the absence of an emitter is shown by exctinction cross section spectra in Fig.~\ref{fig1}(a), for gaps decreasing from 5 nm (blue lines) to 2 nm (green lines) and 0.9 nm (red lines). The latter gap is still outside the tunnelling regime \cite{savage_nat491,scholl_nl13,yan_prl115}, and it is experimentally feasible with unique precision through particle binding to appropriate molecular linkers \cite{chikkaraddy_nat535,taylor_nn5}. As the gap width decreases, interaction between the nanospheres increases and the hybrid dimer bonding plasmon modes drastically redshift \cite{romero_oex14}. Within GNOR (solid lines in all figures), the modes are always blueshifted from the results of LRA (dashed lines in all figures), and significantly broadened, as expected \cite{mortensen_natcom5}. These modes are accompanied by a strong field enhancement and confinement at the gap, even though the near field is significantly reduced in the GNOR description \cite{raza_ol40}, and combine the behaviours of a good cavity and a good antenna, thus becoming promising candidates for enhancing emission \cite{sun_apl98}. This expectation is indeed verified in Fig.~\ref{fig1}(b), where fluorescence enhancement spectra in the presence of the plasmonic cavity are plotted for all gap widths of Fig.~\ref{fig1}(a). As discussed in the Introduction, the hybrid plasmon modes are no longer dark, fluorescence is largely enhanced even at a few Angstroms away from the metal surface, and emission peaks can be immediately associated to the corresponding far-field resonances of the dimer. Within LRA, a fluorescence enhancement of the order of $10^{6}$ is achieved, a value among the highest ever reported for emitter-plasmon systems \cite{rose_nl14,puchkova_nl15}. However, the GNOR corrections show that this enhancement is actually reduced by nearly two orders of magnitude, and emission at the wavelength of higher-order modes is nearly completely damped by the additional loss mechanisms \cite{uskov_plas9,raza_natcom6}. This behaviour is in good agreement with the findings of Larkin \emph{et al.}, who showed through random-phase-approximation studies that the description of an emitter in close proximity to a flat metal surface requires inclusion of spatial dispersion in the metal dielectric function, and that Landau damping produces a dramatic modification of the nonradiative decay rate \cite{larkin_prb69}.

To gain further insight into the mechanisms governing the reduction of the emitted signal predicted by GNOR, we decompose $\gamma_{\mathrm{em}}$ into its two independent components [Figs.~\ref{fig1}(c)-(d)], i.e. emitter excitation by the external field, described by $\gamma_{\mathrm{exc}}$, and emission of energy which can be either radiated or absorbed by the metal, described by $q$. For both processes, it is evident that Landau damping, asdescribed by the GNOR model, leads to a strong decrease of the relevant electric fields [shown in the insets, at the resonant wavelength, within both LRA (right-hand contours) and GNOR (left-hand contours) for each process], thus leading to the nonclassical $\gamma_{\mathrm{em}}$ decrease of Fig.~\ref{fig1}(b). It is also interesting to note that as the gap decreases, absorptive losses and charge screening tend to dominate in the GNOR model, leading to a drastic reduction in the quantum yield not predicted by LRA, which cannot be fully compensated by the increase in the excitation rate. A saturation of the enhancement and an optimum gap width is therefore anticipated, a behaviour not observed within LRA which predicts a divergent fluorescence enhancement \cite{kongsuwan_arxiv}. We verified this tendency by considering even smaller gaps (results not shown here), for which additional suppression mechanisms due to electron spill-out should eventually become important \cite{zhu_natcom7,yan_prl115}, increasing nonradiative losses even further.

\begin{figure}[h]
\centerline{\includegraphics*[width=1\columnwidth]{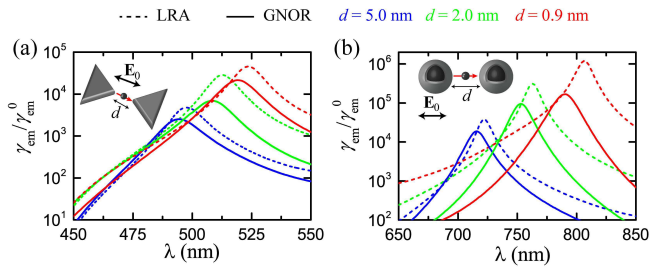}}
\caption{(a) Fluorescence enhancement spectra for a dipole emitter placed at the middle of a bowtie antenna gap, with its dipole moment oscillating parallel to the antenna axis, as shown in the schematics. The bowtie antenna consists of two Ag triangles of height 40 nm and thickness 10 nm. (b) Fluorescence enhancement spectra for a dipole emitter placed at the middle of the gap formed by two SiO$_{2}$/Ag nanoshells ($R_{1} =$ 18 nm, $R =$ 20 nm), with its dipole moment oscillating parallel to the dimer axis, as shown in the schematics. In all cases blue, green, and red lines correspond to $d =$ 5, 2, and 0.9 nm respectively. Solid and dashed lines represent calculations with the GNOR and LRA models, respectively.}\label{fig2}
\end{figure}

While the enhancement achieved with an Ag nanosphere dimer in Fig.~\ref{fig1} is already extremely large, and the influence on nonlocality on it has been clearly demonstrated, it is useful to explore other geometries of particular interest. A prototypical architecture for field-enhancement related applications is the bowtie antenna \cite{kinkhabwala_natphot3,fischer_oex16,dodson_jpcl4}, which owes its popularity mainly to the extreme focusing achievable at the gap between its two narrow tips. In Fig.~\ref{fig2}(a) we place an emitter at the gap of such a bowtie antenna formed by two isosceles Ag trianges of height 40 nm and thickness 10 nm (with slightly rounded edges), a size which allows direct comparison with the nanosphere dimers of Fig.~\ref{fig1}. The fluorescence peak associated with the main bowtie mode for an incident plane wave polarised along its axis exceeds $10^{4}$ within LRA, and in this case it is only slightly reduced when GNOR is considered. This is because the effective interaction area between the two triangles is smaller and nonlocal effects are dominant mainly in the small region around the tips. This less intense interaction between the two nanotriangles accounts also for the smaller, as compared to Fig.~\ref{fig1}, redshift of the plasmon modes as the gap decreases. Nevertheless, despite the better focusing achieved at the tip, the total enhancement is smaller than in the case of nanosphere dimers of Fig.~\ref{fig1}(b), because the size of the bowtie antenna considered here does not ensure a good radiative coupling with the environment, and the quantum yield is relatively small, $q \leq 0.3$ at all wavelengths. In order to improve the situation, more elongated or thicker triangles could be considered as the antenna constituents, leading however to structures in which placing a single emitter exactly at the gap middle is challenging, and large distributions of molecules (with all possible orientations) should be considered instead \cite{kinkhabwala_natphot3}.

Another possibility is to replace the homogeneous Ag nanospheres with Ag nanoshells. Individual thin nanoshells provide much higher enhancements that the corresponding homogeneous spheres, and at the same time allow more flexibility in shifting the plasmon modes to match the emitter wavelength \cite{tserkezis_nscale8}. In Fig.~\ref{fig2}(b) we simulate a dimer consisting of two SiO$_{2}$/Ag nanoshells (core radius $R_{1} =$ 18 nm, total radius $R =$ 20 nm), with SiO$_{2}$ described by a dielectric constant $\varepsilon_{1} = 2.13$. The two main features observed are the dramatic redshift of the modes, due to the combined effect of plasmon hybridisation in the individual nanoshells \cite{prodan_sci302} and interaction between them, and an increase in fluorescence enhancement as compared to the homogeneous nanosphere dimer. Nonlocal corrections within GNOR, which are expected to be important both because of the narrow gaps but also due to the small thickness of the shell, induce again both a peak blueshift and a decrease in the maximum enhancement, which is nevertheless still of the order of $10^{5}$, and in total higher than $10^{3}$ for a wide range of wavelengths. These values indicate that metallic nanoshell dimers are among the most promising plasmonic architectures for such applications.

\begin{figure}[h]
\centerline{\includegraphics*[width=1\columnwidth]{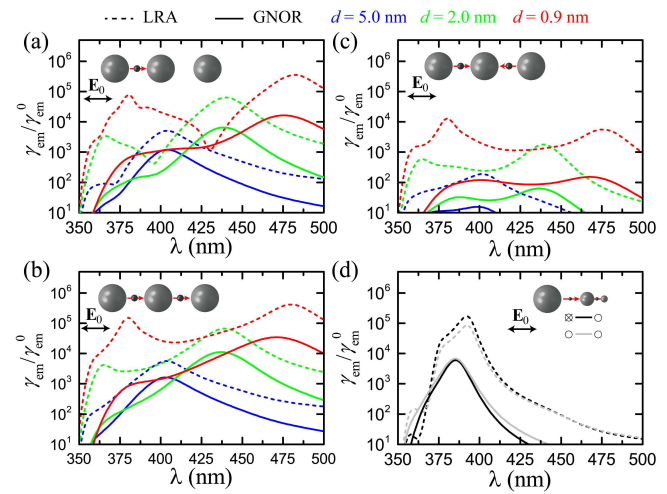}}
\caption{Fluorescence enhancement spectra for dipole emitters placed at the middle of gaps formed by Ag nanosphere trimers, with their dipole moments oscillating parallel to the trimer axis, as shown in the schematics. (a) A single emitter placed at one of the gaps in a trimer of $R =$20 nm spheres. (b) Two emitters placed at the two gaps of the trimer of (a), oscillating in phase and (c) with opposite phases. In all cases blue, green, and red lines correspond to $d =$ 5, 2, and 0.9 nm respectively. Solid and dashed lines represent calculations with the GNOR and LRA models, respectively. (d) A self-similar nanolens consisting of three Ag nanospheres with decreasing radii equal to 20 nm, 8 nm, and 3.2 nm, and accordingly decreasing gaps equal to 2 nm and 0.8 nm (following a geometric progression with common ratio 2.5). Black lines denote a single emitter at the smallest gap, while grey lines denote emitters placed at both gaps and oscillating in phase, as shown in the schematics.}\label{fig3}
\end{figure}

Longer chains can also be created in the same way dimers are fabricated with precise control of the interparticle gaps \cite{taylor_nn5}. Such chains are characterised by long-wavelength strongly radiative modes and high field enhancement \cite{esteban_lang28,tserkezis_ppsc31}, and are therefore well-suited for studying near-field related phenomena \cite{yraola_oex23} and especially emission from molecules at the gaps. In Fig.~\ref{fig3}(a) we consider a chain of three identical Ag nanospheres ($R =$ 20 nm), with a single emitter placed at one of the two gaps (due to the symmetry of the structure, the choice of the gap does not affect the results). Interestingly, the $\gamma_{\mathrm{em}}$ spectra are characterised by a dip in the enhancement, in the wavelength region between the collective chain mode (at about 480 nm for the $d =$ 0.9 nm case) and the higher-order hybrid modes (around 360-380 nm in all cases), which can be understood in view of the dramatic reduction of the scattering cross section (and thus the quantum yield) between the modes (far-field spectra not shown here). Additional damping within the GNOR model significantly smooths many of the distinct features of LRA, leading to more flat spectra, and values reduced again by one order of magnitude.

Placing just one emitter at one of the gaps might be hard to achieve experimentally, and dipoles at both gaps should also be considered. This situation can be achieved for example if organic molecules are coupled to the linkers before particle aggregation \cite{taylor_nn5}. Such a configuration is explored in Fig.~\ref{fig3}(b) for two emitters with parallel dipole moments, where the total excitation rate is defined as the average excitation rate of the two dipoles. Their collective emission increases the total quantum yield, leading to larger $\gamma_{\mathrm{em}}$ values and smoother spectra. Nevertheless, the maximum enhancement is in fact not much improved as compared to Fig.~\ref{fig3}(a). Furthermore, the two emitters could be oscillating out of phase, cancelling each other out in the far field, as shown in Fig.~\ref{fig3}(c), where for the largest of the gaps considered (5 nm), and within GNOR, an almost negligible enhancement is obtained, even on resonance (blue solid lines).

Finally, another interesting architecture based on metallic nanospheres is the self-similar plasmonic lens \cite{li_prl91,almpanis_ol37}, in which nanosphere sizes and interparticle distances gradually decrease following a common geometrical progression (starting from $R = $ 20 nm, $d =$ 2 nm and decreasing with a 2.5 ratio here), leading to a cascaded field enhancement at the smallest gap. Placing the emitter at this gap should then lead to an important increase in the excitation rate, and possibly also in fluorescence enhancement. However, despite these expectations, it was recently shown that while this kind of structure leads to a stronger confinement of the field, in a region of just a few nm, the maximum field value is in fact smaller than in a corresponding dimer of the largest sphere \cite{pellegrini_jpcc120}. Our study shows that, for fluorescence applications, an additional factor limiting the efficiency of self-similar lenses is the fact that the small nanospheres required are always almost completely absorptive, thus reducing the total quantum yield and the signal observed in the far field. In addition, introducing very different nanosphere sizes also leads to a different nonlocal plasmon blueshift for each particle, which means that in practice not all three spheres will be resonant at the same time. The influence of all these limitations can indeed be observed by comparing Fig.~\ref{fig3}(d), where we show fluorescence enhancement spectra for emitters placed at just the smallest gap (black lines) or at both gaps (grey lines), to Fig.~\ref{fig1}(b). 

\begin{figure}[h]
\centerline{\includegraphics*[width=1\columnwidth]{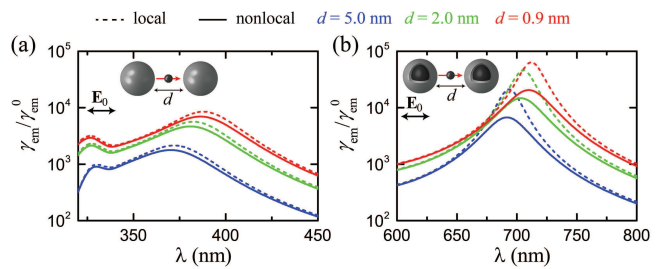}}
\caption{Analytic calculations of fluorescence enhancement spectra for a dipole emitter placed at the middle of the gap between (a) two Ag nanospheres ($R =$ 20nm), and (b) two SiO$_{2}$/Ag nanoshells ($R_{1}$ = 18 nm, $R =$ 20 nm), with its dipole moment oscillating parallel to the dimer axis, as shown in the schematics. Blue, green, and red lines correspond to $d =$ 5, 2, and 0.9 nm respectively. Solid and dashed lines represent calculations within the nonlocal and standard (local) coupled-mode model, respectively.}\label{fig4}
\end{figure}

Having explored a large diversity of geometries, it is useful to adopt analytic models which describe emission enhancement with a simple calculation and can therefore facilitate the design of optimised architectures without the necessity of detailed simulations. The model described in Sec.~\ref{Sec:methods} is applied in Fig.~\ref{fig4}(a) to the Ag nanosphere dimers of Fig.~\ref{fig1}(b), assuming that the original radiative efficiency $\eta_{rad}$ of the emitter is zero \cite{sun_apl98}. It is immediately clear that the general trends predicted by our simulations in Fig.~\ref{fig1}(b) are well reproduced, both in the local and in the nonlocal case, although quantitative differences can be observed, especially concerning the position of the resonances. The coupled-mode model does not fully capture the strength of the interaction between the two particles because, despite its detailed coupling descriptions, it still assimilates the spheres to electric dipoles, thus leading to smaller modal redshifts and lower enhancement values. A similar behaviour is observed for the case of a nanoshell dimer [Fig.~\ref{fig4}(b)], where again the reduction of fluorescence enhancement within nonlocal theory is well reproduced by the analytic solution. It is also worth noting that all the corrections introduced in the nonlocal adaptation of the coupled-mode model enter in terms describing individual particles, and the model fails to accurately capture the tendency of enhancement saturation with decreasing gap. Nevertheless, it still provides efficient, intuitive guidelines for the design of emitter-plasmon cavities for a wide range of interparticle gaps, especially when its nonlocal extension is adopted. Further decreasing the gap towards the Angstrom regime exceeds the limits of applicability of such a model, and requires more elaborate theories \cite{yan_prl115}.

\section{Conclusions}\label{Sec:conclusion}

In summary, we have explored fluorescence enhancement spectra for several plasmonic architectures typically encountered in literature, and showed that simple dimers of homogeneous spheres or nanoshells offer the easiest route to extremely large fluorescence enhancements, relaxing the necessity to resort to more complex geometries. By applying the GNOR model, which efficiently describes both nonlocal screening and surface-enhanced Landau damping in the constituent particles, we showed the existence of an additional fundamental limitation to fluorescence enhancement, which brings theoretical predictions closer to experimental measurements. The underlying enhancement mechanisms were analysed in view of an analytic model for the emission enhancement, which we extended here to include the nonlocal modal shifts and plasmon damping. Our results are expected to facilitate both the design of architectures suitable for fluorescence enhancement, and the interpretation of experiments focused either on emission control or on nonclassical theories for plasmonics.

\begin{acknowledgements}
C.~T. was supported by funding from the People Programme (Marie Curie Actions) of the European Union's Seventh Framework Programme (FP7/2007-2013) under REA grant agreement number 609405 (COFUNDPostdocDTU).
We gratefully acknowledge support from the Villum Foundation via the VKR Centre of Excellence NATEC-II and from the Danish Council for Independent Research (FNU 1323-00087). Center for Nanostructured Graphene (CNG) was financed by the Danish National Research Council (DNRF103).
\end{acknowledgements}

\end{document}